\documentclass[12pt]{article}
\usepackage{enumerate}
\usepackage{amsmath,amsthm}
\usepackage{amssymb, latexsym}
\usepackage[mathscr]{euscript}
\usepackage{color}
\usepackage{array}
\usepackage{flafter}
\usepackage{layout}
\usepackage{graphicx}
\usepackage{epstopdf}
\usepackage[ansinew]{inputenc}
\usepackage{xfrac}
\usepackage{algorithm}
\usepackage{authblk}

\providecommand{\keywords}[1]{\textbf{\textit{Index terms---}} #1}

\theoremstyle{plain}

  \newtheorem{lema}{Lemma}

\theoremstyle{definition}
  \newtheorem{defi}{Definition}
\newtheorem{ej}{Example}

\theoremstyle{remark}

\author[1]{Daniel Andr\'es D\'{\i}az--Pach\'on \thanks{\texttt{Ddiaz3@miami.edu}}}
\author[2]{Robert J. Marks II \thanks{\texttt{ Robert Marks@Baylor.edu}}}
\affil[1]{Division of Biostatistics - University of Miami, Miami, FL}
\affil[2]{Department of Engineering - Baylor University, Waco, TX}

\title{Generalized active information: extensions to unbounded domains}
\date{\today}

\begin{document}

\maketitle

\begin{abstract}
In the last three decades, several measures of complexity have been proposed. Up to this point, most of such measures have only been developed for finite spaces. In these scenarios the baseline distribution is uniform. This makes sense because, among other things, the uniform distribution is the measure of maximum entropy over the relevant space. Active information traditionally assumes a finite interval universe of  discourse but can be extended to other cases where maximum entropy is defined. Illustrating this is the purpose of this paper. Disequilibrium from maximum entropy, measured as active information, can be evaluated from baselines with unbounded support. 
\end{abstract}


\section{Introduction}

%\begin{quote}\begin{center}
%\textit{``Nature abhors a vacuum because she is deeply in love with maximum entropy.''}
%\end{center} \end{quote}
%Aristotle's \textit{horror vacui} postulate here \cite{Ross1963}, the first four words, has proven to be only the iceberg's tip insofar as the ubiquity of maximum entropy, as revealed in statistical mechanics \cite{Cover2012plus,Jaynes1957a,Jaynes1957b}, information theory \cite{Cover2012}, and estimation \cite{Burg1968,Burg1972,Cover2012plus,Kasner2001,Marple1987}. Nature's love of maximum entropy and, more generally, equilibria, helps complete the picture.

Conventional thermodynamic maximum entropy brings to mind the diffusion of an aerosol to be uniformly distributed throughout a closed room. Similarly, Shannon's maximum entropy occurs when the probabilities of an outcome of a fixed number of events are evenly distributed and equal. Maximum entropy and probability as intertwined concepts goes as far back as Jacob Bernoulli, who stated in  1713 \cite{Bernoulli1713}:
\begin{quote}\begin{center}
``... in the absence of any prior knowledge, we must
assume that the events ... have
equal probability.''
\end{center} \end{quote}
Bernoulli's \textit{Principle of Insufficient Reason} (PrOIR) \cite{Papoulis1991} states that, in the absence of further knowledge, the wisest assumption we can do on the distribution of a finite set of points is that its elements are all equiprobable. Bernoulli's idea is so intuitive that it is difficult to dispute. Attempts to do so  have been repudiated \cite{Marks2016}. Interestingly, several developments in science and mathematics during the twentieth century have confirmed its validity. For example:

\begin{enumerate}
        \item {\em Information theory} tells us that equiprobability is the distribution with maximum entropy (maxent) \cite{Jaynes1957a, Jaynes1957b}.
        \item {\em Learning and optimization} tell us that no search can do better uniformly on average than a blind search (i.e., an equiprobable search), which is the average of all existing distributions of a finite sample space \cite{WolpertMacready1995,WolpertMacready1997}.
        \item {\em Bayesian inference} uses equiprobability because it constitutes a noninformative prior over a finite sample space \cite{BoosStefanski2013} (although, of course, other non-informative priors are possible).
\end{enumerate}

The PrOIR is based on absence of further knowledge. We can generalize it to the principle of maximum entropy (maxent), which can be more powerfully asserted ``for the positive reason that [the maxent probability distribution] is uniquely determined as the one which is maximally noncommittal with respect to missing information'' \cite{Jaynes1957a}. A distribution with maximum entropy is such that it is the least biased estimate possible on some given knowledge, albeit partial (see, e.g., \cite{Conrad2004, Jaynes1957b, ParkBera2009}). Such partial knowledge usually takes the form of moments restrictions on a distribution living in a given space.  Table \ref{maxent} below presents the maxent distributions under some moments restrictions over the most relevant spaces \cite{Cover2012}.

	\begin{table}[h!]
		\centering
		\begin{tabular}{| c | c | c|}
			\hline
			Space & Restrictions & Distributions\\
			\hline
			Finite & None & Equiprobability \\
			Finite interval $[a,b]$ & None & $\mathcal U(a,b)$ \\
			$\mathbb Z^+$  & $\textbf EX = \mu$ & Geom($1/\mu$)\\
			$\mathbb R^+$ & $\textbf EX = \mu$ & Exp($1/\mu$)\\
			$\mathbb R$ & $\textbf EX=\mu$; $\textbf E(X-\mu)^2= \sigma^2$ & $\mathcal N(\mu,\sigma^2)$\\
			\hline
		\end{tabular}
		\caption{Maximum entropy distributions over some relevant spaces. $\mathbb Z^+ = \{1,2,\cdots \}$ and $\mathcal N(\mu,\sigma^2)$ is normal with mean $\mu$ and variance $\sigma^2$.
        }\label{maxent}
	\end{table}

In terms of active information, i.e. the information gap between two different distributions, where usually one of the two distributions is taken as a baseline, one may choose the appropriate maxent distribution of Table 1 as this baseline. This change of perspective from the PrOIR to the more general maxent viewpoint serves multiple purposes. First, it answers an old criticism by H\"{a}ggstr\"{o}m on Dembski's preliminary work. H\"{a}ggstr\"{o}m claimed that ``there is absolutely no a priori reason to expect that the `blind forces of nature' should produce a fitness landscape distributed [uniformly]'' \cite{Haggstrom2005}, and repeated his criticism somewhat less rancorously in \cite{Haggstrom2007}. The truth is that we expect a maxent distribution in every aspect of nature, and an explanation is required when some natural process does not follow it.  Moreover, from a subjective point of view, maxent is the safest and most conservative assumption to make, since any other option will introduce selection bias that has not been accounted for. Either way, either in the ontological reality of nature or in our epistemological apprehension of such reality, departures from maxent cannot be explained away. It is not that out-of-equilibrium explanations are not allowed, it is that they must be accounted for. Thus, criticisms to active information, if they aspire to be successful, must be grounded on something else than the baseline distribution being of maxent.

Second, it liberates active information from having to deal with criticisms to the PrOIR \cite{Kasner2001,Ulrych1976}. Many criticisms have been repudiated \cite{Fisher1922,Marks2013,Marks2016}.  For example, Keynes devoted a whole chapter in his \textit{Treatise on Probability} to debunk it \cite{Keynes1921}. Keynes' notions have been addressed by Marks {\em et al.} \cite{Marks2016}.

Third, maxent invites a generalization of active information to other type of spaces beyond the finite one considered until now. With this change, all of the previous results for active information continue being true under the umbrella of a more universal maxent principle. We can evaluate active information under non-compact spaces and justify the use of some endogenous distributions once we have acquired relevant knowledge. Accordingly, we propose to expand the theory of active information using maxent distributions as baselines.

The manifestation of maxent can be viewed thermodynamically. 
\begin{ej}
The second law of thermodynamics famously states that the entropy of a gas in a closed room reaches maximum entropy. Maximum entropy is also manifest in other domains as documented in Table 1.
\end{ej}

\begin{ej}
Barometric pressure is measured from the surface of the earth to space so is an example from Table 1 of maximum entropy distribution of Exp$(1/\mu)$. When the temperature lapse rate is zero, the equation for percent pressure decrease from sealevel is
    $$ \frac{ P (h) }{P_0}= \exp  \left( \frac{-g M h}{RT} \right)  $$
where
    \begin{itemize}
    \item $P(h)$ = pressure at elevation $h$,
    \item $P_0 =$ static pressure at sealevel,
    \item $T = $ temperature in degrees Kelvin,
    \item $h=$ height above sea level,
    \item $R=$ universal gas constant: 8.3144598 J/(mol·K),
    \item $g=$ gravitational acceleration: 9.80665 m/s$^2$, and
    \item $M=$ molar mass of Earth's air: 0.0289644 kg/mol.
    \end{itemize}
\end{ej}

\begin{ej}
The Maxwell-Boltzmann distribution describes the velocity in three dimensions by the vector $\vec{v}=[v_x \; v_y \; v_z]'$ of ideal gas particles as

    $$ f(\vec{v}) = \left( \frac{m}{2\pi kT} \right)^{3/2}e^{-\frac{m\vec v'\vec v}{2kT}}, $$
where $\vec v'$ denotes transposition of the vector and

    \begin{itemize}
    \item $m = $ particle mass,
    \item $T=$ temperature,
    \item $k=$ Boltzman's constant: $1.38064852 \times 10^{-23}$ m$^2$ kg s$^{-2}$ K$^{-1}$
    \end{itemize}

The velocity in each dimension is  in $\mathbb R$ and, in accordance with Table 1, the corresponding maxent displayed in the Maxwell-Boltzmann distribution is normal.
\end{ej}

\begin{ej}
Bernoulli's PrOIR relates to the success of a single trial, $p$. Repeating Bernoulli trials using the same null information assumption of a Bernoulli trial until the first success is, as shown in Table 1, Geom($p$) on
$\mathbb Z^+$ and is, as an extension of a single Bernoulli trial, therefore maxent.
\end{ej}

%Although not universal, maxent is ubiquitous in applications and nature when there are no assumptions to be applied to a problem.

\section{Generalization}
\label{ChristIsRizen0420}

Equilibria in search and learning \cite{Ewert2013a,Marks2016,Mitchell1990,Schaffer1994} as dictated by conservation of information and popularized by the No Free Lunch Theorems \cite{DembskiMarks2009b,Dembski2013,WolpertMacready1995,WolpertMacready1997} dictates that, predicated on maximum Shannon entropy, no search algorithm will outperform any other on average. Active information was introduced to measure the deviation from equilibria, thereby providing a quantification of the information infused in the algorithm to make it work better than average.

Active information can be viewed as a generalized instantiation of {\em anomaly detection} \cite{Allen2012,Chandola2009,Esponda2004,Meneganti1998} otherwise known as {\em novelty filtering}
\cite{Guttormsson1999,Marsland2003,Streifel1996,Thompson2001}. The status quo of probabilistic uniformity is set and any significant deviation is flagged as novel. The degree of deviation from normalcy is measured by the active information. Anomaly detection typically requires training data to establish the quiescence of normalcy. Such is not the case with active information. Normalcy, rather, is defined by conservation of information which, in turn, defines the operational equilibrium of the non-informed search. We will show that this same approach can be applied elsewhere where maximum entropy defined over different domains and under different conditions uniquely establishes a performance equilibria. Monta\~{n}ez has nicely extended the analysis to the information capacity of search spaces and machine learning in general \cite{Montanez2010,Montanez2011,Montanez2013a,Montanez2013b,Montanez2017a,Montanez2017b}.

To derive active information, let $\Omega$ be a \textit{search space} (for instance, one of those considered in Table \ref{maxent}) and $T \subset \Omega$ be a target, where $T$ is a measurable set. Find the maxent distribution $\phi$ for $\Omega$, subject to whatever is known. The \textit{endogenous} information is then defined as
	\begin{align}\label{EndoInfo}
		I_\Omega = -\log \phi (T).
	\end{align}
The units are dependent on the base of the log. When base 2, the information in in bits. Base e gives units of nats and base 10 Hartleys. The endogenous information represents the inherent difficulty  of the problem to reach the target $T$. Any additional knowledge might modify the search generating a new distribution $\varphi$ that will assign probability $\varphi(T)$ of reaching the target $T$. The information required to reach the target when applying this knowledge under $\varphi$ is then \textit{exogenous} and defined as
	\begin{align}\label{ExoInfo}
		I_1 = -\log \varphi(T).
	\end{align}
The difference between (\ref{EndoInfo}) and (\ref{ExoInfo}) naturally defines what is called the \textit{active information} at the target $T$:
	\begin{align}\label{ActiveInfo}
		I_+(\varphi | \phi)(T) = I_\Omega  - I_1 = \log \frac{\varphi(T)}{\phi (T)}.
	\end{align}

Unless otherwise stated, all logarithms are taken to be base 2, so that information is measured in bits. Out of Equation (\ref{ActiveInfo}) the following properties emerge.
%\begin{lema}\label{AIPosit}
%	$I_+(\varphi(T) | \psi(T)) \geq 0$ if and only if $\varphi(T) \geq \psi(T)$.
%\end{lema}
%
%	
%	Lemma \ref{AIPosit} makes clear that the active information can be positive, negative, or zero. It all depends on the knowledge incorporated by the exogenous information. A positive active information means that the exogenous probability is incorporating some knowledge (e.g., the structure of the search space or the target location) that makes the target $T$ more probable than the endogenous one.
\begin{enumerate}
\item A search is improved iff the assisted search has a higher probability that the unassisted search. In other words the active information is positive, i.e. $I_+(\varphi(T) | \phi(T)) \geq 0$ iff $\varphi(T) \geq \phi(T)$. %The corresponding active information is positive in such cases.
\item A negative active information means that the search induced by the exogenous probability is deleterious e.g., the information about the location of $T$ was not accurate. This occurs when $\varphi(T) < \phi(T)$. The active information can approach $-\infty$ as $\varphi (T) \rightarrow 0$.
\item The maximum active information occurs when the probability of success in the alternative search equals one, i.e. $\varphi (T)=1$. Then $I_+ = I_\Omega $ and all available information has been extracted from the search space.
\item Finally, a null active information means that the exogenous information did not contribute anything new in order to reach $T$. This occurs when there is no improvement and $\varphi(T) = \phi(T)$ and the exogenous and endogenous information are identical.
\end{enumerate}
Knowledge cannot always be translated into the search. An interesting example in which additional knowledge about the structure of the search space does not alter the chances of reaching the target can be seen in the next example:
	
\begin{ej}\label{Equip}
	Let $\Omega = \{\omega_1, \omega_2, \ldots, \omega_n\}$ be the initial search space and let $T \subset \Omega$ be the target. Thus the endogenous distribution $\phi$ is given by the equiprobability of all singletons, so that $\phi (T) = |T|/ |\Omega|$. Although equiprobability is sometimes referred to as a discrete uniform distribution, in our context equiprobability is not uniformity, since a uniform r.v. must live in a metric space with some distance metric $d$, and the elements of $\Omega$ must satisfy that $d(\omega_i, \omega_{i-1}) = d(\omega_j, \omega _{j-1})$, for $2\leq i,j \leq n$ (see, e.g., \cite{Dembski1990}). Suppose further that we learn that $\omega_i - \omega_{i-1} = 1$, for $2 \leq i \leq n$. Thus $\phi$ can be replaced by a uniform distribution $\textbf U$. The uniform distribution incorporates the knowledge about the equidistance between points in the space but such knowledge is not detected by active information since $I_+ = \log \textbf U(T)/\phi (T) = 0$.
\end{ej}

The situation in Example \ref{Equip} occurs because information is defined in terms of the \textit{probability} of events, but not on the events as such. Also, notice that active information \textit{might be} introduced when we gain knowledge on the structure of the search (sample) space or on the location of the target. Example \ref{Equip} deals with the first case: active information that \textit{could be} added because we learned something about the space. Possible practical implications of this situation might be seen, for example, when comparing two different categorical r.v.'s over the same space (e.g., from nominal to ordinal, as when we jump from a space in which the only thing we knew about it was its cardinality but then learned that its elements are ordered)  ---or even a categorical r.v. against a uniform one over the same space  provided both r.v.'s assign equal probability to each singleton (e.g., from nominal to uniform, when we learn that the elements of the space are equidistant). Thus, although a nonzero active information implies a change of distribution, not all acquired knowledge about the search space implies a nonzero active information. 

\begin{defi}[Stochastic domination]\label{StochDom}
	Let $\phi$ and $\varphi$ be two (cumulative probability) distributions over $\Omega$. $\phi$ is stochastically dominated by $\varphi$ if $\phi(x) \geq \varphi(x)$ for all $x$.
\end{defi}

\begin{ej}
	Let $\Omega = \{1, 2, 3\}$. And consider two r.v.'s $X$ and $X'$ on $\Omega$, with mass functions $f_X$ and $f_{X'}$ given in Table \ref{dominus}. Calling $F_X$ and $F_{X'}$ the respective distributions of $X$ and $X'$, then $F_X(x)$ is stochastically dominated by $F_{X'}(x)$.
	\begin{table}[h!]
		\centering
		\begin{tabular}{c| c  c  c}
				 & 1 & 2 & 3\\
			\hline
			$f_X$ & 2/3 & 1/6 & 1/6 \\
			$f_{X'}$ & 1/3 & 1/3 & 1/3
		\end{tabular}
		\caption{$F_X(x) \geq F_{X'}(x)$ for all $x\in \Omega$}\label{dominus}
	\end{table}
\end{ej}

\begin{figure}
                \begin{center}
                \includegraphics[width=.67\textwidth]{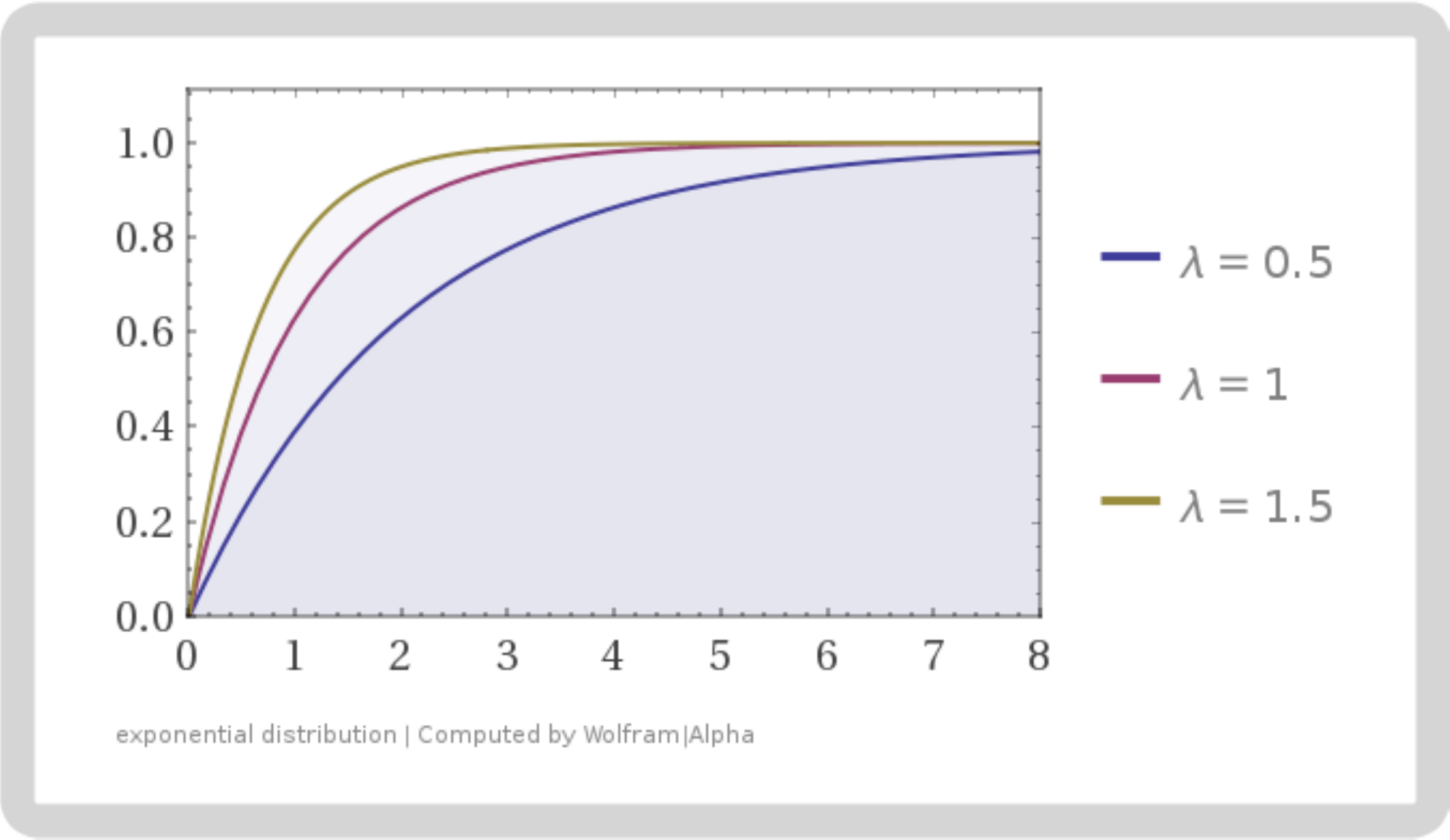} %
                \caption{Three CDF's of the exponential r.v.'s with intensity parameters 0.5, 1, and 1.5. Clearly the one with parameter 1.5 is dominated by the one with parameter 1, which in turn is dominated by the one with parameter 0.5. }
               \label{Exps}
               \end{center}
\end{figure}

\begin{ej}
	Let $X$, $Y$, and $Z$ be exponential r.v.'s with intensity parameters $\lambda=1/\mu=0.5, 1, 1.5$ respectively, and distributions $F_X$, $F_Y$, $F_Z$, respectively. Then $F_Z$ is stochastically dominated by $F_Y$, and $F_Y$ in turn is stochastically dominated by $F_X$. See Figure \ref{Exps}.
\end{ej}
%\begin{figure}[h!]
%  \centering
%    \includegraphics[width= 8 cm]{Exps}
%  \caption{Three CDF's of the exponential r.v.'s with parameters 0.5, 1, and 1.5. Clearly the one with parameter 1.5 is dominated by the one with parameter 1, which in turn is dominated by the one with parameter 0.5.}
%  \label{Exps}
%\end{figure}

\begin{lema}\label{Dominance}
	Let $\phi$ and $\varphi$ as in Definition \ref{StochDom}.  $\varphi$ is stochastically dominated by $\phi$ if and only if $I_+(\varphi | \phi)((-\infty,x]) \geq 0$ for all $x$.
\end{lema}

Thus, whenever $\varphi$ is stochastically dominated by $\phi$, any target of the form $T = (-\infty, x]$ has non-negative active information, and negative active information for any target of the form $T' = (x, \infty)$.

\subsection{Active information as a complexity measure}

Active information can also be seen as a statistical complexity measure. Feldman and Crutchfield argue that at least three things are needed to build a statistical complexity measure: (i) it needs to vanish on the extremes, (ii) it needs to have a clear interpretation, and (iii) it needs to have a specified use \cite{FeldamCrutchfield1998}. Active information, $I_+ = \log \varphi(T)/\phi(T)$, as a statistical measure of complexity \textit{per event} satisfies these requirements:
	
\begin{itemize}		
		\item[(i)] The active information is well defined. It is 0 when the endogenous probability equals the exogenous one ($\phi(T) = \varphi(T)$.) On the other hand, when 						$\varphi(T)=1$ we say that the search is perfect and $I_+ = I_\Omega $, which means that all information available in the problem was extracted in the modified search 				for the target.
		\item[(ii)] The interpretation of $I_+$ is evident from its definition and the four properties listed in Section \ref{ChristIsRizen0420}. It measures how many bits of difference there are 				between the endogenous and the exogenous information at the event $T$. The active information is a measure of the degree to which a search has been assisted to succeed.
		\item[(iii)] When Wolpert and Macready popularized the No Free Lunch Theorems (NFLT), they established the impossibility of one search algorithm to outperform any 				other without additional knowledge \cite{WolpertMacready1995, WolpertMacready1997}. That is, in absence of additional information, no algorithm should do 					better, on average, in finding a target than blind search. In their words, ``the performance of any two algorithms... is, on average, identical'' 								\cite{WolpertMacready1995}. In reaching a target, however, search algorithms usually do better than blind chance. Why is this so? Again, using the words of 					Wolpert and Macready, it is because of the direct input of information, by ``incorporating problem-specific knowledge into the behavior of the [optimization or 					search] algorithm'' \cite{WolpertMacready1997}. Active information determines the information gap between the search of a target by pure chance and the input of 				an expert/dumb programmer.	
\end{itemize}

The usual way to proceed with measures of complexity on a bounded domain is to account for some kind of difference between equilibrium and non-equilibrium distributions, i.e. equiprobability versus non-equiprobability, for finite settings. For instance, L\'opez-Ruiz et al.\ define \textit{disequilibrium}  (i.e., the sum of the quadratic Euclidean distance of the probability of atomic events), $D_E$, as part of their complexity measure \cite{LopezManciniCalbet1995}; Martin et al.\ replace the Euclidean distance by Wootters' \cite{Wootters1981} to obtain their disequilibirum $D_W$, while keeping the difference between equilibrium and non-equilibrium distributions \cite{MartinPlastinoRosso2003}. And finally \textit{active information} is the difference of the information for an event under equilibrium and non-equilibrium \cite{DembskiMarks2009a, DembskiMarks2009b, Ewert2012, Ewert2013b,Marks2010, Montanez2010}. In fact, just as the informational entropy is the average of the self-information, the Kullback--Leibler distance (\cite{Cover2012}, chap. 12), referenced to the maximum entropy distribution, is the average of the active information over all the elements of a finite alphabet.

\subsection{Conclusions}

We have proposed extension of active information for domains other than those confined to a finite interval.
%When confined to the interval $(0,\infty)$, the continuous maximum entropy
Maximum entropy (maxent) is defined on domains other than a finite interval. These maxent manifestations are evidenced in both applications and nature. Using these generalized maxent measures, active information formerly associated with finite domain distributions, can be extended. The extension conforms to criteria for measuring statistical complexity.  

Active information has recently been expanded as a multidimensional mode hunting tool \cite{DiazEtAl2019}. Briefly speaking, in a finite space every deviation from equiprobability constitutes a local mode. Thus in every event of positive active information there is a local mode. In a subsequent paper, the generalized measure is applied specifically to the Wright-Fisher model of population genetics \cite{DiazEtAl2020}.

\section{Acknowledgements}

We thank the reviewers for their valuable insights. This article is much better because of their comments.

\end{document}